\documentclass[aps,prb,superscriptaddress, twocolumn]{revtex4}
\usepackage[dvipdfm]{graphicx}

\begin{document}

\title{Absence of the 1/8 -anomaly in oxygen isotope effect on YBa$_{2}$Cu$_{3}$O$_{y}$ }

\author{K.~Kamiya}
\affiliation{Department of Physics, Osaka University, Osaka 560-0043, Japan.}

\author{T.~Masui}
\affiliation{Department of Physics, Kinki University, Osaka 577-8502 Japan.}

\author{S.~Tajima}
\affiliation{Department of Physics, Osaka University, Osaka 560-0043, Japan.}

\author{H.~Bando}
\affiliation{National Institute of Advanced Industrial Science and Technology (AIST), Ibaraki 305-8568, Japan.}

\author{Y.~Aiura}
\affiliation{National Institute of Advanced Industrial Science and Technology (AIST), Ibaraki 305-8568, Japan.}

\setcounter{tocdepth}{3}
\begin{abstract}
We have prepared oxygen isotope exchanged crystals of impurity-free YBCO with various oxygen concentents, and examined pure doping ($p$) dependance of isotope effect on superconducting transition temperature. With decreasing oxygen contents, the isotope exponent $\alpha$ monotonously increases without any anomaly around $p = 1/8$. The monotonous increase in $\alpha$ indicates that phonons are involved in the mechanism which causes the monotonous $T_c$ suppression with underdoping.

\end{abstract}

\maketitle
~~ The isotope exchange (IE) experiments played a crucial role in establishing the BCS mechanism for superconductivity \cite{Reynolds1950}. Also in the case of the high $T_c$ cuprate superconductors, a lot of efforts have been devoted on oxygen IE experiments \cite{review}. Nevertheless, they could not draw a clear conclusion about the phonon contribution to high $T_c$ superconductivity. First the almost zero IE effect was observed in the 90 K superconductor YBa$_{2}$Cu$_{3}$O$_{y}$, which gave a negative proof against the phonon-mediated pairing mechanism\cite{Batlogg1987}. However, the succeeding experiments showed a finite or even large IE effect on $T_c$, depending on the materials and compositions \cite{review}.  The IE effects on London penetration depth\cite{Hofer2000, Khasanov2004a, Khasanov2004} and the photoemission dispersion curve\cite{Iwasawa2008,Gweon2004} suggested some relation of phonons with superconductivity but in a manner different from the BCS type.

~~The most pronounced contribution of phonons to the electronic state in the cuprates is observed in the so-called stripe order of charge and spin\cite{Tranquada1995}. When $T_c$ is suppressed near the doping level $p = 1/8$ in La$_{2-x}$Sr$_{x}$CuO$_{4}$ (LSCO) or $p = 0.16$ in La$_{1.8-x}$Eu$_{0.2}$Sr$_{x}$CuO$_{4}$, the oxygen IE effect is strongly enhanced\cite{Crawford1990a,Zhao1995,Franck1993a,Sasagawa2005}, which is attributed to the stripe order. In ref.13, it was suggested that the phonon contributes to a de-pairing mechanism rather than a pairing mechanism, based on the idea that the stripe order competes with superconductivity. On the other hand, there are theoretical models in which the stripes play a positive role in superconductivity\cite{Carlson2000, Hirsch2002}.
 
~~Recently a charge density wave (CDW) was reported in underdoped YBa$_{2}$Cu$_{3}$O$_{y}$ (YBCO).  \cite{Chang2012,Ghiringhelli2012} Although the observed charge order is not of a stripe-type in LSCO, it is interesting to examine an IE effect on it. So far, there has been little systematic study on the carrier doping dependence of IE effect in YBCO. This is partly because of the difficulty in exchanging $^{16}$O to $^{18}$O with keeping the same oxygen content for underdoped samples.

~~Instead of changing oxygen content, the substitution of La for Ba or Pr for Y can reduce the carrier doping level. The IE experiments were reported for fully oxygenated Y(Ba, La)$_{2}$Cu$_{3}$O$_{y}$\cite{Bornemann1991} and (Y,Pr)Ba$_{2}$Cu$_{3}$O$_{y}$\cite{Soerensen1995, Fraser1991}. The authors found that the isotope exponent $\alpha$ increases as $T_c$ decreases with increasing La or Pr content, and eventually exceeds the BCS value $\alpha = 0.5$, while it is nearly zero at the optimally and over-doped YBCO \cite{Soerensen1995, Fraser1991, Bornemann1991} (Here $\alpha$ is defined as $\alpha = d (\ln T_{c}) / d (\ln M_{i}) $, where $M_{i}$ is isotope mass.)  In these studies it is not obvious whether the carrier doping level is systematically changed by La or Pr substitution or not, because there are additional factors for $T_c$ suppression such as the effect of Pr 4f orbital\cite{Radousky1992} and the disorders due to La/Ba substitution\cite{Wu1999}.   Therefore, in order to see a pure doping-dependence of IE effect, it is necessary to examine the effect for pure YBCO with changing oxygen content.

~~The purpose of the present work is to examine the doping($p$)-dependence of IE effect in YBCO system, and to search for the CDW related feature in the IE effect of this material. Using impurity-free YBCO crystals with various oxygen contents, we carried out the systematic IE experiment over a wide doping range.

~~Single crystals of YBCO were grown by a pulling-up technique\cite{Shiohara1993}. All the pieces used in this study were cut out from the same as-grown crystal. The sample annealed in the optimal condition (500$^\circ$C in oxygen flow) showed a sharp superconducting transition ($\Delta$ $T_c$ = 0.5 K ) at 93.6 K. Isotope exchange and adjustment of oxygen content for a pair of samples (IE and non-IE) were carried out in the following careful procedure. Firstly, in order to reduce oxygen content, the samples were annealed at 800$^\circ$C in N$_{2}$ flow for 3 days. Then the samples were put in a special tube furnace, in which two identical tubes were equiped. The samples were annealed in the furnace in  $^{16}$O- or $^{18}$O-circulating conditions at the temperature appropriate for the expected oxygen content. In annealing, the condition in both tubes were carefully controled at 760 $\pm$ 3 Torr with oxygen flow rate (1.5 $\pm$ 0.2 cc/min). At the end of annealing, we quenched the samples by quickly pushing them out from the furnace. 

~~The superconducting transition was examined by dc magnetization ($M$) measurement with a field of 10 Oe. As an example, $M(T)$ for the samples annealed at 500$^\circ$C and 640$^\circ$C are shown in Fig.1(a) and (b), respectively. In both cases, one can see a sharp and parallel transition at $T_c$ for a pair ($^{16}$O and $^{18}$O) of samples. All the pairs of samples used in this study exhibited a similar sharp and parallel transition, which guarantees the compositional homogeneity.

~~In order to confirm that a pair of samples have the same oxygen content, we measured the lattice constant $c$ that is sensitive to the oxygen content\cite{Fisher1993}. In the insets of Fig.1(a) and (b), the Cu-K$\alpha$ X-ray diffraction peaks for (0 0 13) and (0 0 14) are compared between the $^{16}$O- and $^{18}$O-samples. The diffraction peaks do not shift at all with IE, namely, the $c$-axis lattice parameter is identical for a pair ($^{16}$O and $^{18}$O) of samples, indicating that the oxygen contents are kept the same.

~~The accomplishment of isotope exchange was confirmed by the frequency shift of oxygen vibration mode in Raman scattering spectra as shown in Fig.1(c) and (d). The peaks at 330 cm$^{-1}$, 440 cm$^{-1}$ and 600 cm$^{-1}$ are ascribed to the vibrations of the in-plane, the apical and the CuO-chain oxygen, respectively\cite{Liu1988}. These three peaks show a clear isotope shift from which we can estimate the volume fraction of $^{18}$O exchange as $>$ 90 \%\cite{Altendorf1991}.

~~The dependence of $T_c$ on annealing temperature is plotted in Fig.2. The $T_c$ value was defined by the onset temperature of magnetization transition. In the case of annealing at 500$^\circ$C, the isotope effect on $T_c$ is very small but finite ($\sim$0.3 K), while there is no isotope effect for the annealing at temperatures lower than 500$^\circ$C. In both of the $^{16}$O- and $^{18}$O-sample series, the annealing at 500$^\circ$C gives the highest $T_c$, namely the optimum doping. This is another support that our annealing treatment was successfully equivalent for a pair ($^{16}$O and $^{18}$O) of samples. When the annealing temperature increases, $T_c$ decreases and shows a slight hump around $T_c = 60$ K, as is expected. The $T_c$ difference between a pair monotonically increases with increasing the annealing temperature.

\begin{figure}[h]
\begin{center}
\includegraphics[height=6.8cm]{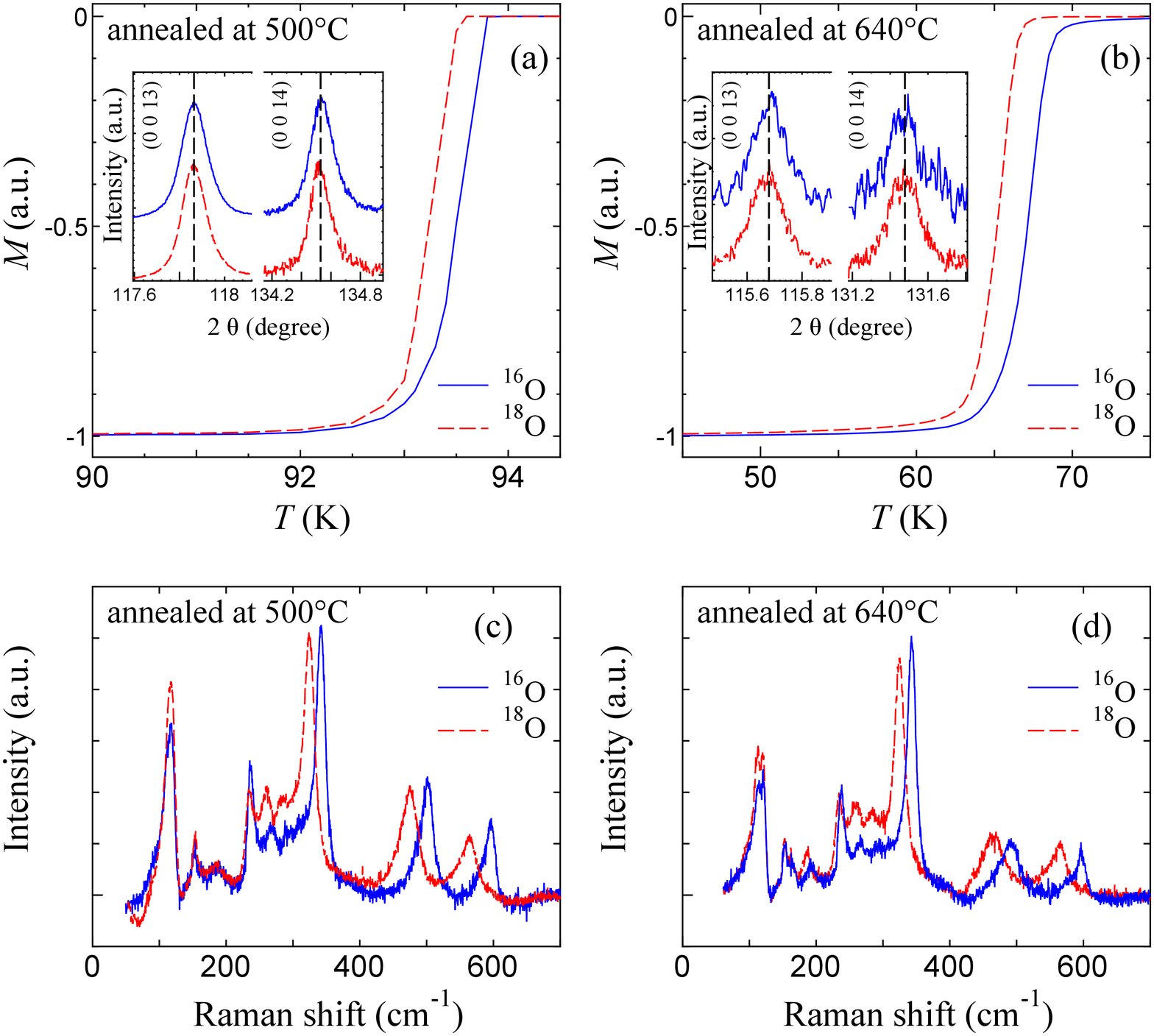}
\end{center}
\caption{(Color online) Temperature dependence of dc magnetization (a, b) and Raman scattering spectra at room temperature (c, d) for the pairs of IE and non-IE samples. The insets of (a) and (b) show the (0 0 \textit{l}) Cu-K$\alpha$ X-ray diffraction peaks (\textit{l}=13 and 14). The
optimally doped samples (a, c) were annealed at 500$^{\circ}$C, 760 Torr, and the underdoped samples (b, d) were at 640$^{\circ}$C, 760 Torr. Note that the temperature scales of (a) and (b) are quite different. }
\label{fig1}
\end{figure}

\begin{figure}[h]
\begin{center}
\includegraphics[height=5cm]{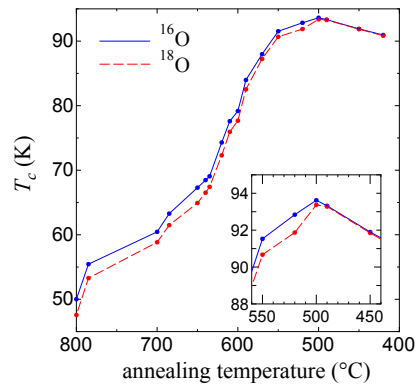}
\label{anneal}
\caption{(Color online) Annealing temperature dependence of $T_c$. The inset shows the expanded figure around the optimum doping.}
\end{center}
\end{figure}

~~The estimated isotope exponents  $\alpha$ were plotted as a function of $T_c$ in Fig.3. The  $\alpha$ monotonically increases from zero to about 0.5 as $T_c$ decreases. This is different from the previous data which showed a sudden decrease of $\alpha$ at $T_c$ = 60 K.\cite{Zech1996} We guess that the origin of the discrepancy may be inhomogeneity of polycrystalline samples in ref.27

~~The $T_c$ dependence of  $\alpha$ in the present study is quite similar to those observed in Pr- or La-substituted YBCO. This implies that the primary reason for $T_c$ suppression is the same, namely, it is caused by the reduction of carrier density in these three cases (La- and Pr-substitution and oxygen reduction).  Although we need to consider the other factors such as the disorders introduced by La-substitution, it turns out to be a minor effect. It is worth noting here that Zn-substitution does not appreciably increase  $\alpha$ until $T_c$ becomes lower than 20 K\cite{Soerensen1995}. A similar observation was reported for Sr-substituted YBCO\cite{Bornemann1991}, where $\alpha$ does not change with $T_c$ suppression by Sr-substitution. These imply that the impurity pair-breaking is not affected by IE.

~~Another important message in Fig.3 is that the observed IE effect is not related to the oxygen at the Cu-O chain site. In the present series of samples, the oxygen occupancy at the chain site decreases with reduction, while in the Pr- and La-substituted YBCO, the chain is almost fully occupied. Therefore, the present result indicates that the isotope effect is connected only with the CuO$_{2}$-planes. This conclusion is in good agreement with the previous report on the site-selective oxygen IE experiment \cite{Zech1994}.

\begin{figure}[h]
\begin{center}
\includegraphics[height=5cm]{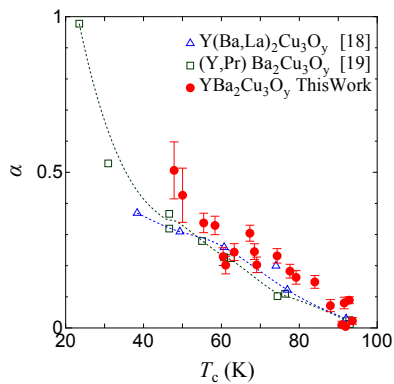}
\end{center}
\caption{(Color online) $T_c$ dependence of $\alpha$ from this work and from the previous report of impurity substituted YBCO system\cite{Bornemann1991, Soerensen1995}. }
\label{alpha}
\end{figure}
 
\begin{figure}[h]
\begin{center}
\includegraphics[height=5cm]{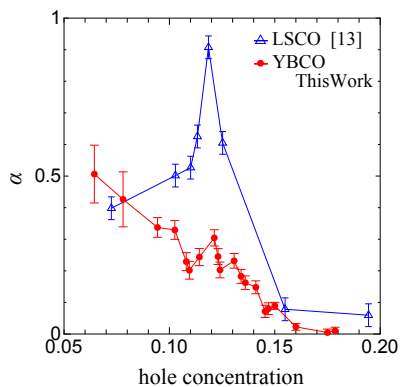}
\end{center}
\caption{(Color online) Carrier concentration dependence of $\alpha$. The result of LSCO system\cite{Sasagawa2005} is also plotted.}
\label{alpha}
\end{figure}

~~Figure 4 demonstrates the carrier doping ($p$) dependence of isotope exponent together with the publishehd result of LSCO\cite{Sasagawa2005}. The values of $p$ were determined from the $T_c$ values of $^{16}$O samples on the basis of the reported data \cite{Liang2006}. In contrast to the result of LSCO with $\alpha$ strongly enhanced near $p = 1/8$, the present result shows a rather monotonic increase in $\alpha$ with reducing $p$. There seems to be a fine structure in $\alpha (p)$ near $p = 1/8$, but it is too weak to be discussed as a meaningful feature. It is concluded that there is no remarkable isotope effect related with the charge density wave or the stripe order in YBCO. Strong magnetic field may be necessary to stabilize the CDW\cite{Chang2012,Ghiringhelli2012}.

~~Finally we discuss the origin of the observed increase in $\alpha$ with underdoping. Since a similar doping dependence of $\alpha$ is observed also in LSCO except for the anomalous increase around $p = 1/8$, we believe that the enhanced isotope effect with reducing $p$ is a common property in high $T_c$ superconducting cuprates.There are two approaches to understand this doping dependence of $\alpha$. One is based on the idea that phonons positively contribute to pairing mechanism. The other is to consider that phonons are involved in de-pairing due to some competing order.

~~There have been several proposals in the first approach\cite{Kresin1997, Phillips1990, Carbotte1990, Tseui1990, Xi1993, Alexandrov1992, Bussmann-Holder2005}.  Kresin${'}$s model of non-adiabatic isotope effect predicts $\alpha \propto \partial T_{{\rm c}}/ \partial n$  and thus non-monotonic change of $\alpha$ with doping \cite{Kresin1997}. But our result in Fig.3 does not agree with this prediction. Some models attribute the unusual doping dependence of $\alpha$ to the van Hove singularity in the electron density of states.\cite{Carbotte1990, Tseui1990, Alexandrov1992} In the polaron mediated mechanism, $\alpha$ decreases with carrier doping($p$) because the polaron binding energy decreases with $p$, whereas the origin of binding energy change is unknown\cite{Xi1993}. Another polaron model\cite{Bussmann-Holder2005} provided a quantitative calculation of $\alpha$($p$) with $p$-independent polaron coupling, based on the band change with $p$. Negative value of $\alpha$ in the overdoped regime is predicted by this model.

~~In the second approach, on the other hand, when $T_c$ decreases with developing a competing order, the IE effect on competing order is expected to become pronounced. The most plausible origin of the competing order is the pseudogap or the Mott related order such as the stripes. Since these competing orders reduce a superconducting condensate via the shrink of the Fermi arc, it is natural that the IE effect was observed in London penetration depth. While some experiments reported the IE effect on the pseudogap\cite{Raffa1998, Khasanov2008a}, however,  the others reported almost no IE effect on it\cite{Williams1998a}.  Since the pseudogap state is not realized as a sharp phase transition but a kind of crossover, it might be difficult to observe the IE effect on the pseudogap. For the elucidation of the high $T_c$ superconductivity mechanism, it is highly important not only to specify the competing order and but also to study the IE effect on it. Moreover, the IE effect in the overdoped regime is crucial to judge which of the first and the second approaches is correct. Some IE effects were reported in the overdoped samples\cite{Bornemann1991a, Bornemann1992} but not well established yet.


~~In summary, we have successfully prepared the oxygen IE crystals of impurity-free YBCO in a wide doping (oxygen content) range. The isotope exponent $\alpha$ is zero in the overdoped samples ($p > 0.16$), while it monotonically increases up to 0.5 with reducing $p$ in the underdoped samples. No remarkable change can be seen near $p = 1/8$ where the anomalous enhancement of $\alpha$ was reported in LSCO and the CDW was observed in YBCO.\\


\end{document}